\documentclass[a4paper,11pt]{article}
\pdfoutput=1 
\usepackage{jheppub}
\usepackage{amsmath}
\usepackage{amsfonts}
\usepackage{amssymb}
\usepackage{slashed}
\usepackage{bm}
\usepackage{graphicx}%
\usepackage[T1]{fontenc} 
\usepackage{mathrsfs}
\usepackage{yfonts}

\DeclareMathAlphabet{\mathpzc}{OT1}{pzc}{m}{it}

\newcommand{\beq}{\begin{equation}} 
\newcommand{\eeq}{\end{equation}} 
\newcommand{\bega}{\begin{eqnarray}} 
\newcommand{\ega}{\end{eqnarray}}

\newcommand{\ie}{i\epsilon}

\newcommand{\dhd}{{\textstyle d}
\lower.03ex\hbox{\kern-0.38em$^{\scriptstyle-}$}\kern-0.05em{}}
\newcommand{\dbar}{{\textstyle \delta}
\lower.03ex\hbox{\kern-0.38em$^{\scriptstyle-}$}\kern-0.05em{}}
\newcommand{\half}{{1\over 2}}

\newcommand{\balef}{{\bar \aleph}}

\newcommand{\baru}{{\bar u}}

\newcommand{\barv}{{\bar v}}

\newcommand{\babeta}{{\bar \beta }}

\newcommand{\belta}{{\bar \delta}}

\newcommand{\bhi}{{\bar \chi}}

\newcommand{\calf}{{\cal F}}

\newcommand{\calk}{{\cal K}}

\newcommand{\calo}{{\cal O}}

\newcommand{\calu}{{\cal U}}

\newcommand{\tilF}{{\tilde F}}

\newcommand{\tamma}{\tilde {\gamma}}

\newcommand{\ve}{\varepsilon}

\newcommand{\frar}{\mathfrak{r}}

\pdfoutput=1
\abstract{High-energy scattering in pQCD in the Regge limit is described by the evolution of Wilson lines
governed by the BK equation \cite{Balitsky:1995ub,Kovchegov:1999yj}. In the leading order, the BK equation is conformally invariant 
and the eigenfunctions of the linearized BFKL equation are powers. It is a common belief that 
at $d\neq 4$ the BFKL equation is useless since unlike $d=4$ case it cannot be solved by usual methods.
However, we demonstrate that at critical Wilson-Fisher point of QCD the relevant part of NLO BK 
restores the conformal invariance so the solutions are again powers. As a check of our approach to high-energy amplitudes at the Wilson-Fisher point, 
we calculate the anomalous dimensions of twist-2 light-ray operators in the Regge limit $j\rightarrow 1$.}
\arxivnumber{}
\affiliation{$^a$ Physics Department, Old Dominion University, Norfolk, VA 23529, USA.\\
$^b$ Thomas Jefferson National Accelerator Facility, Newport News, VA 23606, USA.}
\affiliation{$^c$ Department of Mathematics and Physics, University of Salento, Via Arnesano, Lecce, I-73100, Italy. \\
$^d$ Sezione di Lecce, Istituto Nazionale di Fisica Nucleare, Via Arnesano, Lecce, I-73100, Italy.}
\emailAdd{$^{a,b}$ balitsky@jlab.org,~
$^{c,d}$ chirilli@le.infn.it}

\begin{document}

\title{\boldmath Conformal BK equation at QCD Wilson-Fisher point}
\author{I. Balitsky$^{a,b}$ and G.A. Chirilli$^{c,d}$}
\preprint{JLAB-THY-24-4106}
\maketitle

\flushbottom

\section{Introduction\label{aba:sec1}}
Despite the fact that QCD is  far from being a conformal theory, some aspects of perturbative QCD are very
close to those of a conformal theory. First and foremost, since bare QCD Lagrangian is conformally invariant, 
the leading-log (LL) evolutions in QCD have this property with respect of some subgroup of full $SO(4,2)$.
As an example, one may recall $SL(2,R)$ invariance of the  leading-order evolution of light-ray QCD
operators \cite{Balitsky:1987bk} and $SL(2,C)$ invariance of the BFKL equation \cite{Lipatov:1985uk}. Moreover, 
the common belief is that at higher orders of perturbation theory, the result in ${\cal N}=4$ SYM theory gives the most complicated part of the corresponding 
result in QCD. This so-called maximal transcendentality (MT) feature \cite{Kotikov:2002ab} was checked in several examples. 
For instance, the MT part of three-loop DGLAP kernels is exactly the three-loop result  in ${\cal N}=4$ SYM theory \cite{Kotikov:2010nd}. 
The same property is correct for NLO pomeron intercept \cite{Kotikov:2002ab} and one should expect it to be true for the NNLO 
intercept which unfortunately is not obtained in QCD yet, although the    ${\cal N}=4$ intercept is available \cite{Gromov:2015vua,Caron-Huot:2016tzz,Velizhanin:2015xsa}.

Another example of a conformal theory with perturbation theory ``close'' to that of QCD, is QCD at $d\neq 4$ at the so-called
Wilson-Fisher (WF) point \cite{Wilson:1971dc}  $d=4-2\ve_\ast$ where $\beta$-function vanishes. An example of practical use of this closeness is the restoration
of three-loop result for the GPD evolution in QCD  from that of PDF evolution by a two-loop calculation of Ward identity at
the WF point \cite{Braun:2016qlg,Braun:2017cih,Ji:2023eni}. The goal of this paper is to lay a groundwork for similar analysis for the BFKL/BK equation 
describing high-energy (small-$x$) limit of QCD.

The common belief is that in $d\neq 4$ the BFKL equation is useless since unlike $d=4$ it is not conformally invariant,
 and as a result, cannot  be solved by power-like eigenfunctions.  
 However, we will demonstrate below that at the WF point the conformal 
 invariance of the BFKL equation is restored, at least in the leading order, so the solutions  are powers again. This opens up the 
 whole machinery of conformal field theories (CFTs) like general Conformal Regge theory discussed in Refs. \cite{Costa:2012cb,Costa:2013zra,Costa:2023wfz} 
 and theory of conformal light-ray operators  \cite{Kravchuk:2018htv,Kologlu:2019mfz,Chang:2020qpj,Caron-Huot:2022eqs}.
 
 As an application, we use the conformal BK equation at WF point  to find anomalous dimensions of twist-2 light-ray operators of spin $j$
 in the ``BFKL limit'' $j\rightarrow 1$.  Usually the extraction of $j\rightarrow 1$ limit of anomalous dimensions is done using
 the BFKL kernel in the momentum space \cite{Jaroszewicz:1982gr,Fadin:1998py,Ciafaloni:1998gs},  but in a conformal theory there is an alternative way. Since BFKL equation
 gives the amplitude in the Regge limit and anomalous dimensions are relevant for the DGLAP evolution of LR operators,
 one can consider a correlation function (CF) of 4 scalar currents in the double ``Regge+light-cone'' limit. On one side, one can first take Regge limit of that 
  CF which is given 
 by conformal Regge formula \cite{Cornalba:2007fs} (see below) and  take a light-cone (LC)  limit of that formula afterwords. 
 On the other hand, we can start with an expansion 
 in light-ray (LR) operators. Since LR operators
 are an analytic continuation of local operators (see the discussion in Ref. \cite{Kravchuk:2018htv}) one can use an extension of Polyakov three-point formula
 and get the light-cone expansion of the same CF with explicit dependence on anomalous dimensions.
 After that we can take the small-$x$ $j\rightarrow 1$ limit and compare this ``LC+Regge'' limit to ``Regge+LC'' representation discussed above.  As was mentioned, the former representation involves anomalous dimensions of the leading-twist  LR operators while the latter is governed by the BFKL pomeron. The comparison gives us an opportunity to relate anomalous dimension of LR operator  to BFKL
 intercept.
 The result is  as expected:  the pomeron intercept at $d=4-2\ve_\ast$ is different from intercept at $d=4$, but the anomalous
 dimensions (in the $\overline{MS}$ scheme) are the same.
 
 The paper is organized as follows: in Sect. 2 we derive the conformal LO  BK equation at the WF point using NLO BK in $d=4$.  In Sect. \ref{sec:RLC}
 we derive the ``Regge+LC'' representation discussed above while Sect. \ref{sec:LCR} is devoted to ``LC+Regge'' one. In Sect. \ref{sec:andim} we compare the
 two representations and get the anomalous dimensions of LR operators in the $j\rightarrow 1$ limit while Sect. 4 is a ``conclusions+outlook'' section. 
 The necessary d=4 NLO BFKL kernel and eigenvalues are listed in the Appendix \ref{app:nloeigens} and the formal definition 
 of gluon LR operators is presented in Appendix \ref{app:lrays}.
 
 \section{BK equation at the WF point}
 We consider QCD at non-integer $d=4-2\ve_\ast$ such that the $\beta$-function vanishes
\beq
 {1\over g}\beta(g)~=~-\ve-\babeta(a)~=~-\ve-(b_0a+b_1a^2+...)~=~0,
 ~~~~~~a\equiv{\alpha_s\over 4\pi}
\eeq
To be on the safe side, one should consider QCD at small $\ve$ near the Banks-Zaks point \cite{Banks:1981nn} where $n_f$ is such that $\babeta(a)~=~0$. At small $\ve$ we are still in the conformal window. Since we are interested in the first few orders in perturbative expansion,  after calculation we can analytically continue our formulas to $n_f$ outside of the conformal window.
 
Thus, we consider QCD at  the critical Wilson-Fisher point $\ve_\ast$ such that
\beq
\ve_\ast=-ab_0-a^2b_1,~~~~b_0={11\over 3}N_c-{2\over 3}n_f,~~~b_1={2\over 3}\big[17 N_c^2-5n_fN_c-3c_Fn_f\big]
\label{WFpoint}
\eeq
With out NLO accuracy we need only the $b_0$ part of Eq. (\ref{WFpoint}), the $b_1$ term will be important 
for NNLO calculations.

 The main observation of the paper is the following. 
 The LO BK equation for the rapidity evolution of color dipole ${\rm tr}\{ \hat{U}^\eta_{z_1} \hat{U}^{\dagger\eta}_{z_2}\}$ 
 at arbitrary dimension of the transverse space $d_\perp=d-2$ has the form:
\begin{equation}
 {d\over d\eta}{\rm tr}\{ \hat{U}^\eta_{z_1} \hat{U}^{\dagger\eta}_{z_2}\}
 =~{\alpha_s\over 2\pi^{d_\perp}}{\Gamma^2\big({d_\perp\over 2}\big)\over \mu_{\rm MS}^{d_\perp-2}}
\!\int\!d^{d_\perp}z_3~
K_{\rm LO}^{d_\perp}(z_i)
\Big[{\rm tr}\{U_{z_1}U^\dagger_{z_3}\}{\rm tr}\{U_{z_3}U^\dagger_{z_2}\}-N_c{\rm tr}\{U_{z_1}U^\dagger_{z_2}\}\Big]
\label{devolution}
\end{equation}
where  $\mu_{\rm MS}^2=\mu^2{e^{\gamma_E}\over 4\pi}$  
 \footnote{Throughout the paper,  $\mu$ is the normalization point in 
the $\overline{MS}$ scheme and $\alpha_s\equiv\alpha_s(\mu)$} and
\begin{equation}
K^{d_\perp}_{\rm LO}(z_i)~=~{4\pi^{d_\perp}\over\Gamma^2\big({d_\perp\over 2}\big)}\Big[(z_1|{p_i\over p^2}|z_3)(z_3|{p_i\over p^2}|z_1)
+(z_2|{p_i\over p^2}|z_3)(z_3|{p_i\over p^2}|z_2)
-~2(z_1|{p_i\over p^2}|z_3)(z_3|{p_i\over p^2}|z_2)\Big]
\label{dkernel}
\end{equation}
Here
we use Schwinger's notation for propagators $(x|{1\over p^2}|y)\equiv\int\!{d^np\over(2\pi)^n}e^{-ip(x-y)}$
and the rapidity cutoff  $\eta$ is defined as the logarithm of maximal $p_+$ of the gluons forming Wilson lines 
\beq
U(x_i)~=~{\rm Pexp\Big\{}ig\!\int\! dx_+A_-(x_+,0,x_{i_\perp})\Big\}
\eeq

The formula (\ref{devolution}) is the same as the LO BK kernel in Ref. \cite{Balitsky:1995ub}, with the only difference that
the free propagators (before and after interaction with the shock wave ) are in $d_\perp=2-2\ve$. Now, 
at $d_\perp=2$ we get the familiar conformal dipole kernel \cite{Mueller:1993rr,Mueller:1994jq}
\begin{equation}
4\pi^2\Big[(z_1|{p_i\over p_\perp^2}|z_3)(z_3|{p_i\over p_\perp^2}|z_1)+(z_2|{p_i\over p_\perp^2}|z_3)(z_3|{p_i\over p_\perp^2}|z_2)
-2(z_1|{p_i\over p_\perp^2}|z_3)(z_3|{p_i\over p_\perp^2}|z_2)\Big]
~=~{z_{12}^2\over z_{13}^2 z_{23}^2}
\end{equation}
while at $d_\perp=2-2\ve$ one obtains the kernel
\begin{eqnarray}
&&\hspace{-1mm}
K^{d_\perp}_{\rm LO}(z_i)~=~
{1\over (z_{13}^2)^{d_\perp-1}}+{1\over (z_{23}^2)^{d_\perp-1}}-{2z_{13}\cdot z_{23}\over (z_{13}^2z_{23}^2)^{d_\perp\over 2}}
\end{eqnarray}
which is not conformal so the corresponding LO BFKL equation cannot be solved by powers.

However, at the critical point $\ve_\ast$ one should expand in $\ve$ and $\alpha_s$ simultaneously. 
The expansion of the LO kernel $K^{d_\perp}_{\rm LO}(z_i)$ can be rewritten as
\begin{eqnarray}
&&\hspace{-1mm}
K^{2-2\ve_\ast}_{\rm LO}(z_i)~=~
\Big({z_{12}^2\over z_{13}^2 z_{23}^2}\Big)^{1-\ve_\ast}\big[1+\ve_\ast\ln z_{12}^2\big]+\Big({1\over z_{13}^2}-{1\over z_{23}^2}\Big)
\ln{z_{13}^2\over z_{23}^2}+O(\ve^2)
\end{eqnarray}
 The NLO BK  equation in d=4 QCD reads \cite{Balitsky:2007feb,Balitsky:2009xg}
\begin{eqnarray}
&&\hspace{-2mm}
{d\over d\eta}\big[{\rm tr}\{ \hat{U}^\eta_{z_1} \hat{U}^{\dagger\eta}_{z_2}\}\big]~
\stackrel{d=4}=~{\alpha_s\over 2\pi^2}
\!\int\!d^2z_3~
{z_{12}^2\over z_{13}^2z_{23}^2}\Big\{1
+~{\alpha_s\over 4\pi}b_0\Big[\ln {z_{12}^2\mu^2\over 4}+2\gamma_E
-{z_{13}^2-z_{23}^2\over  z_{12}^2}\ln{z_{13}^2\over z_{23}^2}\Big]
\nonumber\\ 
&&\hspace{-2mm}
+~{\alpha_sN_c\over 4\pi}\Big({67\over 9}-{\pi^2\over 3}-{10n_f\over 9N_c}\Big) \Big\}
\big[{\rm tr}\{\hat{U}^\eta_{z_1}\hat{U}^{\dagger\eta}_{z_3}\}{\rm tr}\{\hat{U}^\eta_{z_3}\hat{U}^{\dagger\eta}_{z_2}\}
-N_c {\rm tr}\{\hat{U}^\eta_{z_1}\hat{U}^{\dagger\eta}_{z_2}\}\big]~+~~{\alpha_s^2N_c\over 16\pi^2}K_{\rm conf}
\nonumber\\
\label{BK4}
\end{eqnarray}
 where $K_{\rm conf}$ is the conformal part  presented in the Appendix \ref{app:nloeigens}.
Consequently,  the NLO BK equation at $d=4-2\ve$ can be written as
\begin{eqnarray}
&&\hspace{-5mm}
{d\over d\eta}{\rm tr}\{ \hat{U}^\eta_{z_1} \hat{U}^{\dagger\eta}_{z_2}\}~
=~{\alpha_s\Gamma(1-\ve)\over 2\pi^{2-\ve}}
\!\int\!d^{2-2\ve}z_3~
\Big({z_{12}^2\over z_{13}^2z_{23}^2}\Big)^{1-\ve}\Big\{1
+~\Big(\ve+{\alpha_s\over 4\pi}b_0\Big)\Big[\ln {z_{12}^2\mu^2\over 4}
\nonumber\\ 
&&\hspace{11mm}
+~2\gamma_E-~{z_{13}^2-z_{23}^2\over  z_{12}^2}\ln{z_{13}^2\over z_{23}^2}\Big]
+{\alpha_sN_c\over 4\pi}\Big({67\over 9}-{\pi^2\over 3}-{10n_f\over 9N_c}\Big) \Big\}
\big[{\rm tr}\{\hat{U}^\eta_{z_1}\hat{U}^{\dagger\eta}_{z_3}{\rm tr}\{\hat{U}^\eta_{z_3}\hat{U}^{\dagger\eta}_{z_2}\}
\nonumber\\
&&\hspace{11mm}-N_c {\rm tr}\{\hat{U}^\eta_{z_1}\hat{U}^{\dagger\eta}_{z_2}\}\big]
+~{\alpha_s^2N_c\over 16\pi^2}K_{\rm conf}~+~O(\alpha_s^3,\alpha_s^2\ve,\alpha_s\ve^2)
\end{eqnarray}
which turns to conformal equation
\begin{eqnarray}
&&\hspace{-5mm}
{d\over d\eta}{\rm tr}\{ \hat{U}^\eta_{z_1} \hat{U}^{\dagger\eta}_{z_2}\}~
=~{\alpha_s\Gamma(1-\ve_\ast)\over 2\pi^{2-\ve_\ast}}
\!\int\!d^{2-2\ve_\ast}z_3~\Big({z_{12}^2\over z_{13}^2z_{23}^2}\Big)^{1-\ve_\ast}
\Big[1+{\alpha_sN_c\over 4\pi}\Big({67\over 9}-{\pi^2\over 3}-{10n_f\over 9N_c}\Big) \Big]
\nonumber\\
&&\hspace{-5mm}
\times~
\big[{\rm tr}\{\hat{U}^\eta_{z_1}\hat{U}^{\dagger\eta}_{z_3}\}{\rm tr}\{\hat{U}^\eta_{z_3}\hat{U}^{\dagger\eta}_{z_2}\}
-~N_c {\rm tr}\{\hat{U}^\eta_{z_1}\hat{U}^{\dagger\eta}_{z_2}\}\big]~+~~{\alpha_s^2N_c\over 16\pi^2}K_{\rm conf}
~+~O(\alpha_s^3,\alpha_s^2\ve_\ast,\alpha_s\ve_\ast^2)
\label{BKinWF}
\end{eqnarray}
at  the critical point $d=4-2\ve_\ast$. 

Thus, the BK equation at $d=4-2\ve_\ast$ is conformally (M\"obius) invariant up to NLO order. \footnote{
For now, we see only the invariance under the inversion in the transverse plane. At $d=4$, it is
 demonstrated in Ref. \cite{Balitsky:2009xg} that  the  LO BK kernel satisfies equations following from standard commutation 
 relations of leading-order SL(2,C) operators.
These relations are not satisfied at $d=4-2\ve_\ast$ which means that the corresponding leading-order SL(2,C) operators
get $O(\alpha_s,\ve)$ corrections similarly to SL(2,R) operators discussed in Refs. \cite{Braun:2013tva,Braun:2016qlg}. The authors hope to return to this matter 
in future publications.
}
It should be noted that this property, albeit expected at the LO, is not a direct consequence of the 
invariance of the WF QCD since rapidity cutoff of Wilson lines violates conformal invariance (even in 
the conformal ${\cal N}=4$ SYM). 
We hope that at the next order in $\alpha_s$  the conformal invariance will survive after appropriate replacement
of the dipole $U_x U^\dagger_y$ by a suitable ``conformal dipole'' similar to the ${\cal N}=4$ SYM case at $d=4$ \cite{Balitsky:2009xg}.

The conformal invariance of BK equation at the QCD critical point opens up the whole machinery of CFT for the analysis
of high-energy amplitudes. For example, applying the logic of this Section in reverse, one could have predicted
the $b_0$ part of NLO BK (\ref{BK4}) coming from the running of the QCD coupling constant. It is worth noting that  for now,  only this $b_0$ part is used in the numerical estimates of the solution of BK equation with running coupling. 

As an another example of using CFT methods for high-energy amplitudes in QCD, we will obtain the anomalous dimensions 
of twist-two gluon operators $F^{-i}\nabla_-^{j-2}F_{~i}^-$ at the ``BFKL point'' $j\rightarrow 1$. We will see that while the pomeron
intercept at $d=4-2\ve_\ast$ is different from that of $d=4$, the anomalous dimensions of these operators in $\overline{MS}$ scheme 
 are the same as discussed in Refs. \cite{Braun:2016qlg,Braun:2017cih,Ji:2023eni}.
 
 Let us start with calculation of the pomeron intercept at $d=4-2\ve_\ast$.  The intercept is the rightmost eigenvalue of 
the BFKL equation which is a linearization of Eq. (\ref{BKinWF}). To find the intercept, it is sufficient to consider 
forward matrix elements of the dipole
$$
{\cal U}(z_1,z_2)\equiv
1-{1\over N_c}{\rm tr}\{\hat{U}(z_{1_\perp})
\hat{U}^{\dagger}(z_{2_\perp})\},~~~~~\langle {\cal U}(z_1,z_2)\rangle~=~\calu(z_{12})
$$
The forward linearized equation (\ref{BKinWF}) takes the form
\begin{eqnarray}
&&\hspace{-1mm}
\frac {d}{d \eta } {\cal U}(z )
~=~{\alpha_sN_{c}\Gamma(1-\ve_\ast)\over 2\pi^{2-\ve_\ast}}
\!\int \! d^{2-2\ve_\ast}z'~\Big[1+
{\alpha_sN_c\over 4\pi}\Big({67\over 9}-{\pi^2\over 3}-{10n_f\over 9N_c}\Big)\Big] \Big({z^2\over
(z-z')^{2}
{z'}^{2}}\Big)^{1-\ve_\ast}
\nonumber\\
&&\hspace{-1mm}
\times~ \big[2\calu(z')-\calu(z)
\big]
+~{\alpha_s^2N_c^2\over 4\pi^3}\!\int\! dz'\calk_{\rm conf}(z,z')~\calu(z')~+~O(\alpha_s^3,\alpha_s^2\ve_\ast,\alpha_s\ve_\ast^2)
\label{nlobkforward}
\end{eqnarray}
where the linearized forward kernel $\calk_{\rm conf}(z,z')$ is presented in the Appendix \ref{app:nloeigens}. 
Since the pomeron have  quantum numbers of the vacuum, it should be related to the 
angle-independent eigenfunctions of this equation, namely  powers 
$\big{(z'}^2/z^2\big)^\xi$. 
Using the integral
\beq
{\Gamma\big({d_\perp\over 2}\big)\over2\pi^{d_\perp\over 2}}\!\int\! d^{d_\perp}z '
\Big({z^2\over {z'}^2(z-z')^2}\big)^{d_\perp\over 2}\Big[2\big({{z'}^2\over z^2}\big)^\xi -1\Big]~
=~\big[\psi\big({d_\perp\over 2}\big)-\psi(\xi)-\psi\big({d_\perp\over 2}-\xi\big)-\gamma_E\big]
\label{integral}
\eeq
we obtain the eigenvalues in the form
\begin{eqnarray}
&&\hspace{-1mm}
\balef(\xi)~=~{\alpha_sN_c\over\pi}\Big[\bhi(\xi)+{\alpha_sN_c\over 4\pi}\belta(\xi)~+~O(\alpha_s^2,\alpha_s\ve_\ast,\ve_\ast^2)\Big]
\nonumber\\
&&\hspace{-1mm}
\bhi(\xi)~=~ \psi(1-\ve_\ast)-\gamma_E-\psi(\xi)-\psi\big(1-\ve_\ast-\xi\big)
\label{alefn}
\end{eqnarray}
where $\psi(x)$ is the logarithmic derivative of the gamma-function ($\psi(1)=-\gamma_E$).
With our accuracy, we need 
 the function $\belta(\xi)$ at $d=4$. It was calculated in Ref.  \cite{Fadin:1998py} and we represent it in the Appendix \ref{app:nloeigens}.
 \footnote{To avoid confusion with the notations in Ref. \cite{Fadin:1998py} we denote our (slightly different)
 functions $\bhi$ and $\belta$ with bars.}
 
It is easy to see that the pomeron intercept is the eigenvalue corresponding to eigenfunction 
with power  $\xi~=~\half-{\ve_\ast\over 2}-i\nu$ (and no angular dependence). Indeed, writing the identity
\begin{eqnarray}
&&\hspace{-1mm}
\calu(|z|)~=~{\Gamma\big({d_\perp\over 2}\big)\over\pi^{d_\perp\over 2}}\!\int d^{d_\perp}z'\!\int\!{d\nu\over 2\pi}({z'}^2)^{-{d_\perp\over 2}-{d_\perp\over 4}+i\nu}
(z^2)^{{d_\perp\over 2}-{d_\perp\over 4}-i\nu}\calu(|z'|),
\end{eqnarray}
taking the derivative with respect to rapidity according to Eq. (\ref{nlobkforward}) and exponentiating, one obtains
\begin{eqnarray}
&&\hspace{-5mm}
\calu(|z|)~=~{\Gamma\big({d_\perp\over 2}\big)\over\pi^{d_\perp\over 2}}
\!\int d^{d_\perp}z'\!\int\!{d\nu\over 2\pi}({z'}^2)^{-{d_\perp\over 2}-{d_\perp\over 4}+i\nu}
(z^2)^{{d_\perp\over 2}-{d_\perp\over 4}-i\nu}~e^{\eta\balef\big({d_\perp\over 4}-i\nu\big)}
\calu(|z'|).
\end{eqnarray}
So the pomeron intercept (at $\nu=0$) is 
${\alpha_sN_c\over\pi}\chi\big(\half-{\ve_\ast\over 2}\big)~=~{\alpha_sN_c\over\pi}\big[4\ln 2-{\pi^2\over 3}\ve_\ast+O(\ve_\ast^2)\big]$.

 \section{Pomeron intercept and anomalous dimensions of the twist-2 light-ray operators}
 
It is well known that the analytic continuation of anomalous dimensions of twist-2 local operators $F_{-i}\nabla_-^{j-2}F_-^{~i}$ 
to $j\simeq 1$ can be obtained from BFKL pomeron intercept.  Since $j=1$ is the unphysical point for  local gluon operators, one 
should consider the generalization of local twist-2 operators to gluon light-ray operators discussed in 
Refs. \cite{Balitsky:1987bk,Balitsky:2013npa}. 
A gluon LR operator is an analytic continuation in spin of a local operator $F_{-i}\nabla_-^{j-2}F_-^{~i}$ , see Appendix \ref{app:lrays} for detailed
definitions. For simplicity, we consider pure gluodynamics where gluon light-ray operators are multiplicatively renormalizable.

As mentioned in the Introduction, usually the relation between Pomeron intercept and anomalous dimensions of twist-2 operators is obtained from the BFKL kernel in the momentum space, but in a conformal theory
it can be done directly in the coordinate space. One way is to consider the correlation function of
two ``Wilson frames" as was done in Ref. \cite{Balitsky:2013npa} for ${\cal N}=4$ SYM. However, this method requires calculating many nontrivial integrals at $d_\perp\neq 2$.  An easier way, which does not require
any computation,  is described in Ref. \cite{Balitsky:2014sqe}. One considers the CF of four scalar operators, 
for example in $d=4-2\ve_\ast$ QCD and takes $\calo(x)=F^a_{\mu\nu}(x)F^{a\mu\nu}(x)$, and get
\begin{eqnarray}
&&\hspace{-8mm}
A(R,r)~=~[x_{12}^2x_{34}^2]^{4-2\ve_\ast+\gamma_\calo}(\mu^2)^{2\gamma_\calo}\langle \calo(x_1) \calo(x_2)\calo(x_3)\calo(x_4) \rangle~
\label{CF}
\end{eqnarray}
where $\gamma_\calo$ is the anomalous dimension of the operator $\calo$. With Regge limit in view, it is convenient to choose two conformal ratios as follows:
\begin{eqnarray}
&&\hspace{-1mm}
R={x_{13}^2x_{24}^2\over x_{12}^2 x_{34}^2},~~~~\frar~=~R^{-1}\Big[1-{x_{14}^2x_{23}^2\over x_{13}^2 x_{24}^2}+{1\over R}\Big]^{-2}
\label{confratios}
\end{eqnarray}
Now, we consider the CF (\ref{CF}) in the double  Regge and light-cone limit.  The Regge limit can be specified as 
\begin{equation}
x_{1+}\rightarrow \rho x_{1+}, x_{2+}\rightarrow \rho x_{2+},~~x_{1-}=x_{2-}=0,~~~x_{3_-}\rightarrow \rho' x_{3_-}, x_{4_-}\rightarrow \rho' x_{4_-}~~,x_{3+}=x_{4+}=0
\label{reggelimit}
\end{equation}
with $\rho\rho'\rightarrow \infty$
and the light-cone limit corresponds to $x_{12_\perp}^2\rightarrow 0$ (we use metric $x^2=2x_+x_--x_\perp^2$) . 
It is convenient to consider ``forward'' correlation function
\footnote{In our formulas throughout the paper $L_+$ and $L_-$ are positive.}
\begin{eqnarray}
&&\hspace{-5mm}
 A(L_+,L_-; x_{1_\perp},x_{2_\perp},x_{3_\perp},x_{4_\perp})
 ~=~(x_{12_\perp}^2x_{34_\perp}^2)^{4-2\ve_\ast+\gamma_\calo}(\mu^2)^{2\gamma_\calo}
 \label{FCF}\\
&&\hspace{-5mm}
\times~
\!\int\! dx_{2_+}dx_{3_-}~\langle\calo(L_++x_{2_+},x_{1_\perp})\calo(x_{2_+},x_{2_\perp})\calo(L_-+x_{4_-},x_{3_\perp})\calo(x_{4_-},x_{4_\perp})\rangle
\nonumber
\end{eqnarray}
in the double  limit:  Regge ($L_+L_-\rightarrow \infty$) plus light-cone ( $x_{12}^2\rightarrow 0$).
 By comparison of two 
 orders: ``Regge+LC'' and ``LC+Regge'' we will be able to relate anomalous dimensions of gluon LR operators 
 in the limit 
 \beq
 j-1=\omega\rightarrow 0,~{\alpha_s\over\omega}\sim 1
 \label{bfklimit}
 \eeq
 to the BFKL pomeron intercept.

\subsection{``Regge+LC' limit of the CF (\ref{FCF}) \label{sec:RLC}}
The 4-point CF (\ref{CF}) is a function of two conformal ratios (\ref{confratios}) which reduce to 
\begin{eqnarray}
&&\hspace{-5mm}
R~=~{x_{13}^2x_{24}^2\over x_{12}^2x_{34}^2}~\rightarrow~{4x_{1+}x_{2+}x_{3_-}x_{4_-}\over x_{12_\perp}^2x_{34_\perp}^2},~~~~~~
\frar~=~R^{-1}\Big[1-{x_{14}^2x_{23}^2\over  x_{13}^2x_{24}^2}+{1\over R}\Big]^{-2}~
\nonumber\\
&&\hspace{-5mm}
\rightarrow~{ x_{12_\perp}^2x_{34_\perp}^2 x_{1+}x_{2+}x_{3_-}x_{4_-}\over 
[x_{24_\perp}^2x_{1+} x_{3_-} +x_{2+}x_{4_-}x_{13_\perp}^2-x_{1+}x_{4_-}x_{23_\perp}^2-x_{2+}x_{3_-}x_{14_\perp}^2]^2}
\label{Rr}
\end{eqnarray}
in the Regge limit (\ref{reggelimit}). 
It is easy to see that under the rescaling (\ref{reggelimit})  $R$ increases with ``energy'' ($\sim\sqrt{\rho\rho'}$) while $\frar$ is energy-independent.
As demonstrated in Ref. \cite{Cornalba:2007fs,Cornalba:2008qf,Costa:2012cb}),  the general formula for 
the Regge limit of a 4-point CF in a conformal theory has the form
\begin{eqnarray}
&&\hspace{-3mm}
A(R,\frar)~
\stackrel{s\sim\rho\rho'\rightarrow\infty}{=}~{i\over 2}\!\int\! d\nu~f_+(\aleph(\alpha_s,\nu))
F(\alpha_s,\nu)
\Omega(\frar,\nu)R^{\aleph(\alpha_s,\nu)/2}
\label{cornalba}
\end{eqnarray}
where $f_+(\omega)={e^{i\pi\omega}-1\over \sin\pi\omega}$ is a signature factor and
\begin{eqnarray}
&&\hspace{-1mm}
\Omega(\frar,\nu)
=~~{2\nu\sinh 2\pi\nu\Gamma\big(2-{d\over 2}+2i\nu\big)\Gamma\big({d\over 2}-1-2i\nu\big)\Gamma(d-2)\over 2^{d-1}\pi^{d+1\over 2}\Gamma\big({d\over 2}-\half\big)}
C^{{d\over 2}-1}_{-{d\over2}+1+2i\nu}\big({1\over 2\sqrt{\frar}}\big)
\label{Omega}
\end{eqnarray}
is a solution of the Laplace equation for $H_{d-1}$ hyperboloid $(\partial^2_{H_{d-1}}+\nu^2+1)\Omega(\frar,\nu)=0$.
The dynamics is described by the pomeron intercept $\aleph(\alpha_s,\nu)$ and the ``pomeron residue''
$F(\alpha_s,\nu)$.  
The formula (\ref{cornalba}) was proved in Ref. \cite{Cornalba:2007fs} (see also Refs. \cite{Cornalba:2008qf,Costa:2012cb}) by considering the 
leading Regge pole in a conformal theory.
Also, it was demonstrated up to the NLO level that the structure (\ref{cornalba}) is reproduced by 
the high-energy OPE in Wilson lines \cite{Balitsky:2009yp}. 

Now let us take the light-cone limit $x_{12}^2=-x_{12_\perp}^2\rightarrow 0$ on top of the Regge limit (\ref{cornalba}). In this limit
\begin{eqnarray}
&&\hspace{-1mm}
R={x_{13}^2x_{24}^2\over x_{12}^2 x_{34}^2}~\rightarrow~{4x_{1_+}x_{2_+}x_{3_-}x_{4_-}\over x_{12_\perp}^2 x_{34_\perp}^2},~~~~~~~
\frar~\rightarrow~{x_{1_+} x_{2_+} x_{3_-} x_{4_-}x_{12_\perp}^2 x_{34_\perp}^2\over 
x_{12_+}^2(x_{3_-}x_{14_\perp}^2-x_{4_-}x_{13_\perp}^2)^2}
\end{eqnarray}
and
\beq
\hspace{-0mm}
\Omega(\frar,\nu)~\stackrel{\frar\rightarrow \infty}
=~{2^{\ve_\ast -2}\nu\sinh2\pi\nu\Gamma(2-2\ve_\ast)\over \pi^{{5\over 2}-\ve_\ast}\Gamma\big({3\over 2}-\ve_\ast\big)\Gamma(1-\ve_\ast)}
\Big[\Gamma(-2i\nu)\Gamma(1-\ve_\ast+2i\nu)~\frar^{\half-{\ve_\ast\over 2}+i\nu}
+{\rm c.c.}\Big]
\label{omega}
\eeq
so Eq. (\ref{FCF}) reduces to
\begin{eqnarray}
&&\hspace{-1mm}
 A(L_+,L_-; x_{i_\perp})
 ~=~
 ~i\!\int\! d\nu~f_+(\aleph(\alpha_s,\nu))
\tilF(\alpha_s,\nu)\!\int\! dx_{2_+}dx_{4_-}
\theta(L_+ +x_{2_+})\theta(-x_{2+})
\nonumber\\
&&\hspace{26mm}
\times~\theta(L_- +x_{4_-})\theta(-x_{4_-})
\Big({(L_+ +x_{2_+})x_{2_+}(L_- +x_{4_-})x_{4_-}\over x_{12_\perp}^2 x_{34_\perp}^2}\Big)^{\aleph(\nu,\alpha_s)\over 2}
\nonumber\\
&&\hspace{26mm}
\times~
\Big({L_+^2[(L_-+x_{4_-})x_{14_\perp}^2-x_{4_-}x_{13_\perp}^2]^2\over (L_+ +x_{2_+}) x_{2_+} (L_- +x_{4_-})x_{4_-}x_{12_\perp}^2 x_{34_\perp}^2}\Big)^{-\half+{\ve_\ast\over 2}-i\nu}
\label{fcf2}
\end{eqnarray}
where the prefactors in Eq. (\ref{omega}) are absorbed in the function $\tilF(\alpha_s,\nu)$. 

The theta-function factors in Eq. (\ref{fcf2}) require some explanation. Qualitatetively, one should put them in since 
Regge limit corresponds to $x_{1+},x_{3_-}>0$ and $x_{2+},x_{4_-}<0$. Quantitatively, it follows from the derivation of Eq. (\ref{cornalba}) by high-energy factorization \cite{Balitsky:2009yp}. In that approach, the amplitude (\ref{FCF}) is a product of two ``impact factors'' - coefficients of high-energy OPE in color dipoles, and the amplitude of dipole-dipole scattering. The ``top'' impact factors is the CF of two currents $\calo(L_++x_{2_+},x_{1_\perp})\calo(x_{2_+},x_{2_\perp})$ in the background of the shock wave positioned at $x_+=0$, so the region $L_++x_{2_+}>0>x_{2+}$ contributes with the weight one (there is an intersection with the shock wave) while regions  $L_++x_{2_+},x_{2+}>0$ and $L_++x_{2_+},x_{2+}<0$ give zero
contribution since there are no interactions with the shock wave in that regions. Similarly, from the consideration of
the ``bottom'' impact factors one sees that only the region $L_-+x_{4_-}>0>x_{4_-}$ gives nonzero contribution. 
\footnote{For  similar but more detailed discussion, see Ref. \cite{Balitsky:2013npa}}

Now, introducing variables $v={|x_{2+}|\over L_+}, ~u={|x_{4_-}|\over L_-}$  we get
\begin{eqnarray}
&&\hspace{-7mm} 
A(L_+,L_-; x_{i_\perp},x_{2_\perp},x_{3_\perp},x_{4_\perp})~ 
=~iL_+L_-
\nonumber\\
&&\hspace{-7mm}
\times~
\!\int_0^1\! dudv\!\int\!{d\nu}F(\nu,\alpha_s) 
\Big({x_{12_\perp}^2x_{34\perp}^2\baru u\barv v\over [x_{13_\perp}^2u+x_{14}^2\baru]^2}\Big)^{\half-{\ve_\ast\over 2}+i\nu}
\Big({L_+^2L_-^2\baru u\barv v\over x_{12_\perp}^2x_{34\perp}^2}\Big)^{\aleph(\nu,\alpha_s)/2}f_+(\aleph(\nu,\alpha_s))   ~ 
\nonumber
\end{eqnarray}
Performing integral over $v$ one obtaines the representation of the CF (\ref{FCF}) in the double 
``Regge+LC'' limit in the form
\begin{eqnarray}
&&\hspace{-1mm} 
A(L_+,L_-; x_{1_\perp},x_{2_\perp},x_{3_\perp},x_{4_\perp})~=~\!\int_{\half-i\infty}^{\half+i\infty}\!{d\xi\over 2\pi}  
~f_+\big(\balef(\xi)\big)~F(\xi,\alpha_s)(L_+L_-)^{1+\aleph(\xi)}
\nonumber\\
&&\hspace{50mm} 
\times~
\!\int_0^1\! du~
{(\baru u)^{1-\ve_\ast-\xi+{\aleph(\xi)\over 2}}(x_{12_\perp}^2x_{34_\perp}^2)^{1-\xi-\ve_\ast-{\balef\over 2}}\over [x_{13_\perp}^2u+x_{14_\perp}^2\baru]^{2-2\xi-2\ve_\ast}}~~~
\label{bfkl}
\end{eqnarray}
where $\xi~=~\half-i\nu-{\ve_\ast\over 2}$, $\balef(\xi,\alpha_s)\equiv\aleph\Big(-i\big(\half-\xi-{\ve_\ast\over 2}\big),\alpha_s\Big)$, and 
\beq
F(\xi,\alpha_s)~=~{\Gamma^2\big(2-\ve_\ast-\xi+{\aleph(\xi)\over 2}\big)\over \Gamma(4-2\ve_\ast-2\xi-\aleph(\xi))}
\tilF\Big(\alpha_s, -i\big(\half-\xi-{\ve_\ast\over 2}\big)\Big)
\eeq

\subsection{``LC+Regge'' limit of the CF (\ref{FCF}) \label{sec:LCR}}
This time we start with the light-cone limit $x_{12_\perp}^2\rightarrow 0$ and perform the  light-cone expansion of two ``top'' operators $\calo(x_1)\calo(x_2)$ in light-ray gluon operators. We will demonstrate that
\begin{eqnarray}
&&\hspace{-1mm}
(x_{12_\perp}^2)^{3-\ve_\ast}(\mu^2x_{12_\perp}^2)^{\gamma_\calo}\!\int\! dx_{2_+}~\calo(L_++x_{2_+},x_{1_\perp})\calo(x_{2_+},x_{2_\perp})
\nonumber\\
&&\hspace{-1mm}
=~\!\int_{{3\over 2}-i\infty}^{{3\over 2}+i\infty}\! {dj\over 2\pi}~d(j,\alpha_s)L_+^{j}
(\mu^{2}x_{12_\perp}^2)^{\gamma_j\over 2}\calf^{(+)}_{j}(x_{1_\perp})~+~O(x_{12_\perp}^2)
\label{lcexp}
\end{eqnarray}
where $\calf^{(+)}_j(x_\perp)$
is a ``forward'' light-ray operator with  spin $j$ defined in Eq. (\ref{nonlocalforward}) in the Appendix \ref{app:lrays}.  After that, we will write down three-point CF of light-ray operator $\calf^{(+)}_j(x_\perp)$ with two ``bottom'' operators $\calo(x_3)\calo(x_4)$ 
and obtain the representation of 4-point CF (\ref{FCF}) as an integral over spin $j$. The Regge limit would correspond to the behavior of 
the integrand around $j=1$. 

It is convenient to start with the three-point CF of the LR operator $\calf^{(+)}_j(x_\perp)$ and two local operators. A textbook formula for CF of three primary local operators
- one with spin $n$ and two scalars - in a conformal theory reads
\beq
\hspace{-0mm}
\langle \calo_{\mu_1...\mu_n}(z_1)\calo(z_2)\calo(z_3)\rangle
\sim~ {[1+(-1)^n] \Big({z_{12}\over z_{12}^2}-{z_{13}\over z_{13}^2}\Big)_{\mu_1}...
 \Big({z_{12}\over z_{12}^2}-{z_{13}\over z_{13}^2}\Big)_{\mu_n}
 \over (z_{12}^2)^{\Delta_1+\Delta_2-\Delta_3\over 2}(z_{13}^2)^{\Delta_1+\Delta_3-\Delta_2\over 2}(z_{23}^2)^{\Delta_2+\Delta_3-\Delta_1\over 2}}
\label{tbookfla}
\eeq
In our case, the conformal gluon operators with spin $n$ are made from Gegenbauer polynomials $C^{5\over 2}_n\big({\stackrel{\leftrightarrow}\nabla / \stackrel{\leftrightarrow}\partial}\big)$ (at the  one-loop level). However, for our purposes we need formula (\ref{tbookfla}) integrated over $z_{1+}$ so the only surviving term in $\calo_G^{(n)}$ will be 
$ F^a_{-i}\nabla_-^{n-2}F_-^{a i}$. 

As explained in Ref. \cite{Kravchuk:2018htv}, to be on the safe side, the analytic continuation in spin 
of 3-point CF (\ref{tbookfla}) should be done for Wightman CFs 
 rather than for time-ordered ones so first we need to split this time-ordered CF into
Wightman ones. Anticipating the next integral over total translation in $x_-$, consider the case $x_{3_-}>x_{4_-}$. 
First, we write down
\begin{eqnarray}
&&\hspace{-1mm}
\langle\calo(x_{3_-},x_{3_\perp})\calo_G^{(n)} (x_{0_+},x_{0_\perp})\calo(x_{4_-},x_{4_\perp})\rangle~
\nonumber\\
&&\hspace{-1mm}
=~\theta(x_{3_-}>0>x_{4_-})\langle \calo(x_{3_-},x_{3_\perp})\calo_G^{(n)} (x_{0_+},x_{0_\perp})\calo(x_{4_-},x_{4_\perp})\rangle^W
\nonumber\\
&&\hspace{-1mm}
+~\theta(x_{3_-}>x_{4_-}>0)\langle\calo(x_{3_-},x_{3_\perp}) \calo(x_{4_-},x_{4_\perp})\calo_G^{(n)} (x_{0_+},x_{0_\perp})\rangle^W
\nonumber\\
&&\hspace{-1mm}
+~\theta(0>x_{3_-}>x_{4_-})\langle\calo_G^{(n)} (x_{0_+},x_{0_\perp})\calo(x_{3_-},x_{3_\perp}) \calo(x_{4_-},x_{4_\perp})\rangle^W
\label{3orders}
\end{eqnarray}
where $\langle ...\rangle^W$ denotes  a vacuum average of operators as they stand (without time ordering). 
First, let us consider the first term in the above equation.
After integration over $x_{0_+}$ we get
\begin{eqnarray}
&&\hspace{-1mm}
(\mu^2)^{\gamma_\calo+{\gamma_n\over 2}}
\!\int\! dx_{0_+}\langle\calo(x_{3_-},x_{3_\perp})\ F^a_{-i}\nabla_-^{n-2}F_-^{a i}  (x_{0_+},x_{0_\perp})\calo(x_{4_-},x_{4_\perp})\rangle^W~\stackrel{x_{3_-}>0>x_{4_-}}=
\label{locintegrated}\\
&&\hspace{-1mm}
=~\int\! dx_{0_+}{c_n[1+(-1)^n]\over (x_{34_\perp}^2)^{3-\ve_\ast+\gamma_\calo-{\gamma_n\over 2}}}
\Big({x_{3_-}\over 2x_{0_+}x_{3_-}+x_{03_\perp}^2+\ie x_{3_-}}-{x_{4_-}\over 2x_{0_+}x_{4_-}+x_{04_\perp}^2-\ie x_{4_-}}\Big)^n
\nonumber\\
&&\hspace{-1mm}
\times~{1\over [(2x_{0_+}x_{3_-}+x_{03_\perp}^2+\ie x_{3_-})(2x_{0_+}x_{4_-}+x_{04_\perp}^2-\ie x_{4_-})]^{1-\ve_\ast+{\gamma_n\over 2}}}
\nonumber\\
&&\hspace{-1mm}
=-\pi i{c_n[1+(-1)^n]\over (x_{34_\perp}^2)^{4-2\ve_\ast+\gamma_\calo}}
{\Gamma(1+2n+\gamma_n-2\ve_\ast)\over\Gamma^2(1+n+\half\gamma_n-\ve_\ast)}
\big({x_{34_\perp}^2\over x_{3_-}|x_{4_-}|}\big)^{1+{\gamma_n\over 2}-\ve_\ast}\Big({x_{03_\perp}^2\over x_{3_-}}-{x_{04_\perp}^2\over x_{4_-}}\Big)^{2\ve_\ast-1-n-\gamma_n}
\nonumber
\end{eqnarray}
Now we can continue analytically in $j$ using Eqs. (\ref{nonlocalforward})
\begin{eqnarray}
&&\hspace{-5mm}
-i\pi(\mu^2)^{\gamma_\calo+{\gamma_j\over 2}}
\langle\calo(x_{3_-},x_{3_\perp})  \calf^{(+)}_j(x_{0_\perp})\calo(x_{4_-},x_{4_\perp})\rangle^W~
\stackrel{x_{3_-}>0>x_{4_-}}=
\label{lr3tochkae}\\
&&\hspace{-3mm}
=
-i\pi{c(j,\alpha_s)[1+e^{i\pi j}]\over (x_{34_\perp}^2)^{4-2\ve_\ast+\gamma_\calo}}
{\Gamma(1+2j+\gamma_j-2\ve_\ast)\over\Gamma^2(1+j+\half\gamma_n-\ve_\ast)}
\Big({x_{34_\perp}^2\over x_{3_-}|x_{4_-}|}\Big)^{1+{\gamma_j\over 2}-\ve_\ast}\Big({x_{03_\perp}^2\over x_{3_-}}+{x_{04_\perp}^2\over |x_{4_-}|}\Big)^{2\ve_\ast-1-j-\gamma_j}
\nonumber
\end{eqnarray}
For now, we have proved this equation only for Wightman CF. However, it is easy to see that two other orderings 
in Eq. (\ref{3orders}) give zero result after integration over $x_{0+}$. Indeed, at $x_{3-}>x_{4-}>0$ all singularities
in the denominators in Eq. (\ref{locintegrated}) are of the form $(x_{0_+}+\ie)$ which gives zero contribution to
the integral.  Similarly, at $0>x_{3-}>x_{4-}$ the singularities in the denominators are $\sim (x_{0_+}-\ie)$ and
the result vanishes. This is in accordance with the general statement in Ref. \cite{Kravchuk:2018htv} that
the CFs
$$
\langle \calf^{(+)}_j(x_{0_\perp})\calo(x_{4_-},x_{4_\perp})\calo(x_{3_-},x_{3_\perp})\rangle^W~=~
\langle \calo(x_{4_-},x_{4_\perp})\calo(x_{3_-},x_{3_\perp})\calf^{(+)}_j(x_{0_\perp})\rangle^W~=~0
$$
because a (light-ray) operator with non-integer spin annihilates the vacuum. Thus, we get for the T-product of operators
\begin{eqnarray}
&&\hspace{-5mm}
-i\pi(\mu^2)^{\gamma_\calo+{\gamma_j\over 2}}
\langle\calo(x_{3_-},x_{3_\perp})  \calf^{(+)}_j(x_{0_\perp})\calo(x_{4_-},x_{4_\perp})\rangle~
\stackrel{x_{3_-}>x_{4_-}}=~-i\pi
{c(j,\alpha_s)[1+e^{i\pi j}]\over (x_{34_\perp}^2)^{4-2\ve_\ast+\gamma_\calo}}
\nonumber\\
&&\hspace{-3mm}
\times~
{\Gamma(1+2j+\gamma_j-2\ve_\ast)\over\Gamma^2(1+j+\half\gamma_n-\ve_\ast)}
\Big({x_{34_\perp}^2\over -x_{3_-}x_{4_-}}\Big)^{1+{\gamma_j\over 2}-\ve_\ast}\Big({x_{03_\perp}^2\over x_{3_-}}-{x_{04_\perp}^2\over x_{4_-}}\Big)^{2\ve_\ast-1-j-\gamma_j}
\theta(x_{3_-})\theta(-x_{4_-})
\nonumber\\
\label{lr3tochka}
\end{eqnarray}

Finally, let us perform an integration over total translation in $x_-$ direction:
\begin{eqnarray}
&&\hspace{-3mm}
(\mu^2)^{\gamma_\calo+{\gamma_j\over 2}}(x_{34_\perp}^2)^{4-2\ve_\ast+\gamma_\calo}\!\int\!dx_{4_-}
\langle\calo(L_-+x_{4_-},x_{3_\perp})  \calf^{(+)}_j(x_{0_\perp})\calo(x_{4_-},x_{4_\perp})\rangle~
\label{LR3tochka1}\\
\nonumber\\
&&\hspace{-3mm}
=~-i\pi L_-^jc(j,\alpha_s)[1+e^{i\pi j}]
{\Gamma(1+2j+\gamma_j-2\ve_\ast)\over\Gamma^2(1+j+{\gamma_j\over 2}-\ve_\ast)}
\!\int_0^1\! du {(x_{34_\perp}^2)^{1+{\gamma_j\over 2}-\ve_\ast}(\baru u)^{j+{\gamma_j\over 2}-\ve_\ast}
\over (x_{03_\perp}^2u+x_{04_\perp}^2\baru)^{1+j+\gamma_j-2\ve_\ast}}
\nonumber
\end{eqnarray}
where $u={|x_{4_-}|\over L_-}$ and $\baru\equiv 1-u$.

 Now we are in a position to  prove expansion in LR operators  (\ref{lcexp}). 
 To this end, first we take Eq. (\ref{LR3tochka1}), make the 
top$\leftrightarrow$bottom replacement $x_{3_\perp}\rightarrow x_{1_\perp}, x_{4_\perp}\rightarrow x_{2_\perp}$, 
$x_{3_-}\rightarrow x_{1_+},x_{4_-}\rightarrow x_{2_+}, x_{0_+}\rightarrow x_{0_-}$ and 
$\calf^{(+)}_j(x_{0_\perp})\rightarrow \calf^{(-)}_j(x_{0_\perp})$ 
\footnote{The explicit form of $\calf^{(-)}_j(x_{0_\perp})$ is written in Eq. (\ref{nonlocalminus}) from 
Appendix \ref{app:lrays}.}
and take the light-cone  limit $x_{12_\perp}\rightarrow 0$. We get
\begin{eqnarray}
&&\hspace{-7mm}
(\mu^2)^{\gamma_\calo+{\gamma_j\over 2}}(x_{12_\perp}^2)^{4-2\ve_\ast+\gamma_\calo}\!\int\!dx_{2_+}
\langle\calo(L_++x_{2_+},x_{1_\perp})  \calf^{(-)}_j(x_{0_\perp})\calo(x_{2_+},x_{2_\perp})\rangle~
\label{LR3tochka2}\\
&&\hspace{-3mm}
\stackrel{x_{12_\perp}\rightarrow ~0}=~-i\pi L_+^j{c(j,\alpha_s)[1+e^{i\pi j}]\over
( 1+2j+\gamma_j-2\ve_\ast)}
{(x_{12_\perp}^2)^{1+{\gamma_j\over 2}-\ve_\ast}\over (x_{01_\perp}^2)^{1+j+\gamma_j-2\ve_\ast}}
\nonumber
\end{eqnarray}

Second, the CF of two LR operators has the form  \cite{Balitsky:2013npa,Balitsky:2018irv}
\begin{equation}
\langle \calf^{(+)}_{j={3\over 2}+i\nu}(x_\perp)\calf^{(-)}_{j'={3\over 2}+i\nu'}(y_\perp)\rangle
={2\pi\delta(\nu-\nu')a(j,\alpha_s)\big(1+e^{i\pi j}\big)\over [(x-y)_\perp^2]^{j+1-2\ve_\ast+\gamma_j}(\mu^2)^{\gamma_j}}
\label{cf2lrse}
\end{equation}
Note that it is essential to have T-product of two LR operators in the l.h.s., for the Wightman product one gets zero as discussed in Ref. \cite{Kravchuk:2018htv}
\footnote{We have checked by explicit calculation in the free massless electrodynamics that non-zero contributions to 
CF in Eq. (\ref{cf2lrse}) come
from operator orderings like $F_{+i}(L_- +x_-)\calf_-^j(y_\perp)F_+^{~i}(x_-)$ with one of the LR operators split into two pieces. One may wonder if there will be problems with split gauge links  for QCD light-rays, 
but the analysis in Ref. \cite{Balitsky:2013npa} shows that there are none, at least in the Regge limit.}.

Now let us expand  the l.h.s. of Eq. (\ref{LR3tochka2}) at $x_{12_\perp}\rightarrow 0$ using Eq. (\ref{lcexp}):
\begin{eqnarray}
&&\hspace{-1mm}
(\mu^2)^{{\gamma_j\over 2}}(x_{12_\perp}^2)^{1-\ve_\ast}\!\int_{{3\over 2}-i\infty}^{{3\over 2}+i\infty}\! {dj'\over 2\pi}~d(j',\alpha_s)L_+^{j'}
(\mu^{2}x_{12_\perp}^2)^{\gamma_{j'}\over 2}\langle \calf^{(+)}_{j'}(x_{1_\perp}) \calf^{(-)}_j(x_{0_\perp})\rangle
\nonumber\\
&&\hspace{-1mm}
=~iL_+^{j}\big(1+e^{i\pi j}\big)a(j,\alpha_s)d(j,\alpha_s)(x_{12_\perp}^2)^{1-\ve_\ast+{\gamma_j\over 2}}
(x_{01_\perp}^2)^{2\ve_\ast-j-1-\gamma_j}
\end{eqnarray}
Comparing this to the r.h.s. of Eq. (\ref{LR3tochka2}), we confirm Eq. (\ref{lcexp}) with 
\beq
d(j,\alpha_s)~=~{c(j,\alpha_s)\over a(j,\alpha_s)}
{-\pi\over 1+2j+\gamma_j-2\ve_\ast}
\eeq
Finally, we combine the light-cone expansion (\ref{lcexp}) with CF (\ref{LR3tochka1}) and obtain
\begin{eqnarray}
&&\hspace{-5mm}
 A(L_+,L_-; x_{1_\perp},x_{2_\perp},x_{3_\perp},x_{4_\perp})
 ~=~[x_{12}^2x_{34}^2]^{4-2\ve_\ast+\gamma_\calo}(\mu^2)^{2\gamma_\calo}
 \label{dglap}\\
&&\hspace{-5mm}
\times~
\!\int\! dx_{2_+}dx_{4_-}~\langle\calo(L_++x_{2_+},x_{1_\perp})\calo(x_{2_+},x_{2_\perp})\calo(L_-+x_{4_-},x_{3_\perp})\calo(x_{4_-},x_{4_\perp})\rangle
\nonumber\\
&&\hspace{-5mm}
=~\!\int_{{3\over 2}-i\infty}^{{3\over 2}+i\infty}{dj\over 2\pi}(1+e^{i\pi j})(L_+L_-)^jF(j,\alpha_s)
\!\int_0^1\! du~{(\baru u)^{j+{\gamma_j\over 2}-\ve_\ast}(x_{12_\perp}^2 x_{34_\perp}^2)^{1-\ve_\ast+{\gamma_j\over 2}}\over (x_{03_\perp}^2\baru +x_{04_\perp}^2u)^{1+j+\gamma_j-2\ve_\ast}}
\nonumber
\end{eqnarray}
where 
\beq
 F(j,\alpha_s)~=~-i\pi c(j,\alpha_s)d(j,\alpha_s)
 {\Gamma(1+2j-\gamma_j+2\ve_\ast)\over\Gamma^2(1+j+{\gamma_j\over 2}-\ve_\ast)}
 \eeq
The  ``Regge limit''  of Eq. (\ref{dglap}) corresponds to the behavior of the integrand around $j=1$ 
 which should be compared to behavior of the integrand of Eq. (\ref{bfkl}) at small $\aleph(\xi)$.
 
 \subsection{Anomalous dimensions at $j\rightarrow 1$ from comparison of the two limits \label{sec:andim}}
 It is easy to see that the integrals (\ref{bfkl}) and (\ref{dglap}) are actually the same if one makes identifications
\beq
 \omega~\equiv~j-1~=~\balef(\xi,\alpha_s),~~~~~\gamma_j(\alpha_s)~=~-2\xi-\balef(\xi)~=~-2\xi-\omega
 \label{identification}
\eeq
To compare with classical result of Refs. \cite{Fadin:1998py,Ciafaloni:1998gs} one should rewrite these two equations (\ref{identification}) in terms of 
$\tamma_w=-{\gamma_j\over 2}$ 
 and combine them as follows
\beq
 \omega~=~\balef\Big(\tamma_\omega-{\omega\over 2}\Big) 
 \label{lipequation}
\eeq
Next, we need to solve this equation at small $\omega$ and $\tamma\simeq0$.  With the NLO accuracy one obtains
\beq
 \omega~\simeq~\balef(\tamma_\omega)-{\omega\over 2}\balef'(\tamma_\omega)
 ~\simeq~\balef(\tamma_\omega)-{1\over 2}\balef(\tamma_\omega)\balef'(\tamma_\omega)
 \label{lipequation1}
\eeq
which can be rewritten as
\beq
 \omega~=~{\alpha_sN_c\over\pi}\Big\{\bhi(\tamma_\omega)+{\alpha_sN_c\over 4\pi}\Big[\belta(\tamma_\omega)-2\bhi(\tamma_\omega)\bhi'(\tamma_\omega)\Big]\Big\}
 \label{lipequation2}
\eeq

In the BFKL limit (\ref{bfklimit}) the anomalous dimensions are represented as a sum of series' in 
${\alpha_s\over\omega}$
\beq
\tamma_\omega~=~\sum a_n\big({\alpha_s\over\omega}\big)^n+\omega\sum b_n\big({\alpha_s\over\omega}\big)^n
+\omega^2\sum c_n\big({\alpha_s\over\omega}\big)^n+...
\label{general}
\eeq
The NLO BFKL result (\ref{lipequation2}) gives an opportunity to get the first two series in this expansion.
 Unfortunately, we can inverse Eq. (\ref{lipequation2}) only perturbatively and get a few $a_n$'s and $b_n$'s.  
 To this end, we expand the r.h.s. of 
 Eq. (\ref{lipequation2})  in powers of $\tamma_\omega$. To compare with  result of Ref. \cite{Fadin:1998py} we
 need three terms of the expansion: $\tamma_\omega^{-2}, \tamma_\omega^{-1}$, and constant. 
 
 The expansion of $\chi(\gamma)$ defined in Eq. (\ref{alefn}) has the form
\beq
\chi(\gamma)~=~ \psi(1-\ve_\ast)-\gamma_E-\psi(\gamma)-\psi(1-\ve_\ast-\gamma)~=~{1\over\gamma}+2\ve_\ast\zeta(3)\gamma+...~=~{1\over\gamma} + O(\gamma)
\label{expxi}
\eeq
because to compare with Ref. \cite{Fadin:1998py}  we neglect terms $\sim\alpha_s^2\gamma$  in the r.h.s. of 
Eq .  (\ref{lipequation2}). The expansion of $\belta(\gamma)-2\bhi(\gamma)\bhi'(\gamma)$ is presented in 
Eq. (\ref{deltaexpansion}) in the Appendix,
and combining it with the above expansion one obtains the equation
\begin{eqnarray}
&&\hspace{-1mm}
\omega ~=~{\alpha_sN_c\over\pi}\Big\{{1\over\tamma_\omega}+{\alpha_sN_c\over 4\pi}\Big[
-{11\over 3}{1\over\tamma_\omega^2}
+~2\zeta(3)-{395\over 27}+{11\over 18}\pi^2+O(\tamma_\omega)\Big]\Big\}
\end{eqnarray}
which is literally equation (23) from Ref. \cite{Fadin:1998py} at $n_f=0$. For completeness, let us present the corresponding anomalous dimension
\beq
\tamma_\omega~=~{\alpha_sN_c\over\omega}+O\big({\alpha_s\over\omega}\big)^4
+\omega\Big[-{11\over 12}{\alpha_sN_c\over\pi\omega}
-~\big({\alpha_sN_c\over\pi\omega}\big)^3{1\over 4}\Big(
{395\over 27}-{11\over 18}\pi^2-2\zeta(3)\Big)+O\big({\alpha_s\over\omega}\big)^4\Big]
\label{lipatovgamma}
\eeq
We see that at the critical point $d=4-2\ve_\ast$ we get the same anomalous dimensions as at $d=4$ in accordance with analysis of Refs. \cite{Braun:2016qlg,Braun:2017cih,Ji:2023eni}. Incidentally, both the corrections proportional to $b_0$ in QCD result and correction $\sim\ve_\ast$ 
do not contribute to Eq. (\ref{lipatovgamma}) so the only difference with QCD calculation of Ref. \cite{Fadin:1998py} 
is the origin of the subtraction $\belta(\tamma_\omega)-2\bhi(\tamma_\omega)\bhi'(\tamma_\omega)$ in Eq. 
(\ref{lipequation2}). In classical momentum-space calculation of Ref. \cite{Fadin:1998py},
this subtraction comes from the change of the energy scale from the symmetric scale $QP$ to
 non-symmetric $Q^2$.  Instead, in the coordinate space it comes directly from the symmetric ``energy scale'' $R$, through 
 the identification  $\gamma_\omega=\xi+{\balef(\xi)\over 2}$ in Eq. (\ref{identification}) coming from the comparison
 of two representations of 4-point CF (\ref{CF}) in the Regge limit.

\section{Conclusions}
The main result of the paper is that high-energy amplitudes at QCD Wilson-Fisher point (\ref{WFpoint}) are determined 
 by the Mobius invariant BFKL/BK equation, at least up to NLO level. It should be emphasized that this result  is not a
trivial consequence of the conformal invariance of WF QCD  because BK equation describes evolution of 
color dipoles with respect to rapidity cutoff, and the rapidity cutoff violates conformal invariance. \footnote{It is safe to say that the rapidity cutoff respecting conformal invariance, if it exists, has not been discovered up to now.}
As we mentioned in the introduction, the fact that high-energy amplitudes of WF QCD are determined by (pertubatively 
calculable) BFKL pomeron opens up a whole machinery of CFTs at high energies. We considered one example: calculation
of anomalous dimensions of twist-two LR operators in the BFKL limit (\ref{bfklimit}). Another possible application worth exploring is the relation between high-energy amplitudes and non-global logs which was used in 
Refs. 
\cite{Caron-Huot:2015bja,Caron-Huot:2016tzz}
to get the NNLO BK kernel in ${\cal N}=4$ SYM \footnote{Simon Caron-Huot, private communication.}.

Of course, the next question is how to relate results obtained at the WF point to real pQCD. In Sect. 2 we explained how to find
terms proportional to $b_0$ in the NLO result by performing the LO calculations at the LO level. One may hope  to find terms proportional to $b_0$ in the NNLO BK
kernel is a similar way: calculate
$O(\epsilon)$ corrections to NLO BK integrals from Ref. \cite{Balitsky:2007feb} and say that 
$O(\alpha_s\epsilon_\ast)$ deviations from the conformal NLO kernel  at the WF point are due 
to $O(\alpha_s^2b_0)$ terms in NNLO BK kernel. The study is in progress.

The authors are indebted to V. Braun, P. Kravchuk, and A. Manashov for valuable discussions. 
One of us (I.B.) is grateful to CERN TH for kind hospitality  while a part of this work was done.
This work of I.B. is supported by DOE contract DE-AC05-06OR23177  and by the grant DE-FG02-97ER41028.
The work of G.A.C. is supported by the grant PRIN 2022BP52A MUR "The Holographic Universe for all Lambdas" Lecce-Naples.

\section{Appendix: NLO BFKL eigenvalues \label{app:nloeigens}}
The conformal part of NLO BK equation for QCD ``composite dipole'' of Ref. \cite{Balitsky:2009xg} has the form
\begin{eqnarray}
&&\hspace{-2mm}
K_{\rm conf}~
=~{\alpha_s^2\over 16\pi^4}
\int \!d^2 z_3d^2 z_4~{z_{12}^2\over z_{13}^2z_{34}^2z_{24}^2}\Big\{2\ln{z_{12}^2z_{34}^2\over z_{14}^2z_{23}^2} +
\Big[1+{z_{12}^2z_{34}^2\over z_{13}^2z_{24}^2-z_{14}^2z_{23}^2}
\Big]
\ln{z_{13}^2z_{24}^2\over z_{14}^2z_{23}^2}
\nonumber\\ 
&&\hspace{12mm}
\times~[2{\rm tr}\{\hat{U}^\eta_{z_1}\hat{U}^{\dagger\eta}_{z_3}\}{\rm tr}\{\hat{U}^\eta_{z_3}\hat{U}^{\dagger\eta}_{z_4}\}{\rm tr}\{\hat{U}^\eta_{z_4}\hat{U}^{\dagger\eta}_{z_2}\}
-{\rm tr}\{\hat{U}^\eta_{z_1}\hat{U}^{\dagger\eta}_{z_3} \hat{U}^\eta_{z_4}U_{z_2}^{\dagger\eta}\hat{U}^\eta_{z_3}\hat{U}^{\dagger\eta}_{z_4}\}
\nonumber\\
&&\hspace{33mm} 
-~{\rm tr}\{\hat{U}^\eta_{z_1}\hat{U}^{\dagger\eta}_{z_4} \hat{U}^\eta_{z_3}
U_{z_2}^{\dagger\eta}\hat{U}^\eta_{z_4}\hat{U}^{\dagger\eta}_{z_3}\}-(z_4\rightarrow z_3)]
\nonumber\\
&&\hspace{-2mm} 
+~{\alpha_s^2\over 16\pi^4}\int \!{d^2 z_3d^2 z_4\over z_{34}^4}
\Big\{-2+{z_{13}^2z_{24}^2+z_{23}^2z_{14}^2- 4z_{12}^2z_{34}^2\over z_{13}^2z_{24}^2-z_{14}^2z_{23}^2}
\ln{z_{13}^2 z_{24}^2\over z_{14}^2z_{23}^2}+~\big(z_3\leftrightarrow z_4\big)\Big\}
\nonumber\\ 
&&\hspace{12mm}
\times~\big[\big({\rm tr}\{\hat{U}^\eta_{z_1}\hat{U}^{\dagger\eta}_{z_3}\}{\rm tr}\{\hat{U}^\eta_{z_3}\hat{U}^{\dagger\eta}_{z_4}\}
{\rm tr}\{\hat{U}^\eta_{z_4}\hat{U}^{\dagger\eta}_{z_2}\}
-{\rm tr}\{\hat{U}^\eta_{z_1}\hat{U}^{\dagger\eta}_{z_3}\hat{U}^\eta_{z_4}
\hat{U}^{\dagger\eta}_{z_2}\hat{U}^\eta_{z_3}\hat{U}^{\dagger\eta}_{z_4}\}
\big)
-(z_4\rightarrow z_3)\big]
\nonumber\\
&&\hspace{-2mm} 
+~{\alpha^2_sn_f \over 2\pi^4}\!\int\!{d^2z_3 d^2z_4\over z_{34}^4}
\Big\{2-{z_{13}^2z_{24}^2+z_{23}^2z_{14}^2- z_{12}^2z_{34}^2\over z_{13}^2z_{24}^2-z_{14}^2z_{23}^2}
\ln{z_{13}^2 z_{24}^2\over z_{14}^2z_{23}^2}\Big\}
\nonumber\\
&&\hspace{33mm} 
\times~{\rm tr}\{t^a\hat{U}^\eta_{z_1}t^b\hat{U}^{\dagger\eta}_{z_2}\}
{\rm tr}\{t^a\hat{U}^\eta_{z_3}t^b(\hat{U}^{\dagger\eta}_{z_4}-\hat{U}^\eta_{z_3})\}
\label{kconf}
\end{eqnarray}
We need also the linearized version of this kernel in the ``forward'' case 
when $\langle\calu(x,y)\rangle=\calu(x-y)$:
\begin{equation}
\calk_{\rm conf}(z,z')~=~{z^2\over{z'}^2}\Big[-{1\over(z-z')^2}\ln^2{z^2\over {z'}^2}+\Phi(z,z')
+6\pi\zeta(3)\delta^{(2)}(z-z')+F(z,z')\Big]
\label{linkernel}
\end{equation}
where $F(z,z')$ and $\Phi(z,z')$ are given by Eqs. (55) and (56) from Ref. \cite{Balitsky:2012bs}. 
\begin{eqnarray}
&&\hspace{-1mm}
\Phi(z,z')~=~{z^2\over {z'}^2}\bigg\{ {(z^2-z{z'}^2)\over (z-z')^2(z+z')^2}
\Big[\ln{z^2\over{z'}^2}\ln{z^2{z'}^2(z-z')^4\over (z^2+{z'}^2)^4}
+2{\rm Li}_2\big(-{{z'}^2\over z^2}\big)
\nonumber\\
&&\hspace{-1mm}
-~2{\rm Li}_2\big(-{z^2\over {z'}^2}\Big)\Big]-\Big(1-{(z^2-{z'}^2)^2\over (z-z')^2(z+z')^2}\Big)\Big[\!\int_0^1-\int_1^\infty\Big]
{du\over (z-z'u)^2}\ln{u^2{z'}^2\over z^2} \bigg\}
\end{eqnarray}
\begin{eqnarray}
&&\hspace{-1mm}
F(z,z')~=~{z^2\over {z'}^2}\bigg\{\Big(1+{n_f\over N_c^3}\Big){3(z,z')^2-2z^2{z'}^2\over16z^2{z'}^2}
\Big({2\over z^2}+{2\over {z'}^2}+{z^2-{z'}^2\over z^2{z'}^2}\Big)\ln{z^2\over {z'}^2}-\Big[3
\\
&&\hspace{-1mm}
+~\Big(1+{n_f\over N_c^3}\Big)\Big(1-{(z^2+{z'}^2)^2\over 8z^2{z'}^2}
+{3z^4+3{z'}^4-2z^2{z'}^2\over 16z^4{z'}^4}(z,z)^2\Big)\Big]\!\int_0^\infty\!dt{1\over z^2+t^2{z'}^2}\ln{1+t\over |1-t|} \bigg\}
\nonumber
\end{eqnarray}
It should be noted that the first term in Eq. (\ref{kconf}) and the first three terms in Eq. (\ref{linkernel}) are
actually the ${\cal N}=4$ result which has maximal transcendentality in comparison to the remainder of the QCD result. 
This is especially clear if one writes down
the corresponding eigenvalues 
\begin{eqnarray}
&&\hspace{-1mm}
{1\over\pi}\!\int\! d^2z'~\big({{z'}^2\over z^2}\big)^{\xi-1}{\ln^2{{z'}^2\over z^2}\over (z'-z)^2}
~=~\chi''(\xi),~~~~\chi''(\xi)+2\chi(\xi)\chi'(\xi)~=~8\zeta(3)+O(\xi)
\nonumber\\
&&\hspace{-1mm}
{1\over\pi}\!\int\! d^2z~\Phi(z,z')\big({{z'}^2\over z^2}\big)^\xi~=~
-4\big[S_{-2,1}(\xi-1)+S_{-2,1}(-\xi)+{5\over 4}\zeta(3)\big]+{\pi^3\over\sin\pi \xi}
\nonumber\\
&&\hspace{-1mm} 
=~{\pi^2\over 3\xi}+4\zeta(3)+O(\xi)
\nonumber\\
&&\hspace{-1mm}
F(\xi)~=~{1\over\pi}\!\int\! d^2z'~F(z,z')\big({{z'}^2\over z^2}\big)^\xi
~=~-\Big[3+\Big(1+{n_f\over N_c^3}\Big)
{2+3\xi(1-\xi)\over (3-2\xi)(1+2\xi)}\Big]
{\pi^2\cos\pi\xi\over(1-2\xi)\sin^2\pi\xi},
\nonumber\\
&&\hspace{-1mm}
=~-{1\over\xi^2}\Big({11\over 3}+{2n_f\over 3N_c^3}\Big)-{1\over\xi}\Big({67\over 9}+{13n_f\over 9N_c^3}\Big)
-{395\over 27}+{11\over 18}\pi^2-{71 n_f\over 27 N_c^3}+{\pi^2n_f\over 9N_c^3}
+O(\xi)
\label{Fi}
\end{eqnarray}
The representation of the r.h.s. of the second Eq. (\ref{Fi}) in terms of harmonic sums was obtained in Ref. \cite{Costa:2012cb}.

Thus,  $\belta(\xi)$, defined by Eq. (\ref{alefn}), has the form
\begin{eqnarray}
&&\hspace{-1mm}
\belta(\xi)-2\bhi(\xi)\bhi'(\xi)~=~\chi(\xi)\Big({67\over 9}-{\pi^2\over 3}-{10n_f\over 9N_c}\Big)-\chi''(\xi)
+F(\xi)+\Phi(\xi)+6\zeta(3)-2\chi(\xi)\chi'(\xi)
\nonumber\\
&&\hspace{-1mm} 
=~-{1\over\xi^2}\Big({11\over 3}+{2n_f\over 3N_c^3}\Big)-{1\over\xi}\Big({10n_f\over 9N_c}+{13n_f\over 9N_c^3}\Big)
+2\zeta(3)-{395\over 27}+{11\over 18}\pi^2-{71 n_f\over 27 N_c^3}+{\pi^2n_f\over 9N_c^3}
\label{deltaexpansion}
\end{eqnarray}
%
\section{Appendix: gluon light-ray operators \label{app:lrays}}
First, we need to define a light-ray operator as an analytic continuation of a local operator with spin $j$.  
As explained in Ref. \cite{Kravchuk:2018htv}, a proper way is to continue in spin 
a non-local object rather than the local one. A suitable non-local object  is a ``light transform'' of a local operator 
- integral of the  local operator along 
the null line with some weight \cite{Kravchuk:2018htv}. In our case, we define ``forward'' local operator  as
\begin{equation}
\calo^{(+)}_j(z_\perp)~=~\int\! dz_+ F^a_{-i}\nabla_-^{j-2}F_-^{ai}(z_+,0,z_\perp)
\label{localforward}
\eeq
which is a specific case of  the light transform, namely the integral over null line with unit weight. The analytic continuation of Eq. (\ref{localforward}) in spin $j$ has the form
\bega
&&\hspace{-1mm}
\calf_j^{(+)}(z_\perp)~=~\int\! dz_+ \!\int\! dz'_+\Big[{i\Gamma(j-1)\over 2\pi(z'_++\ie)^{j-1}}+{i\Gamma(j-1)\over 2\pi(-z'_++\ie)^{j-1}}\Big]
\nonumber\\
&&\hspace{14mm}
\times~F^a_{-i}(z_++z'_+,0,z_\perp)[z_++z'_+ +z_\perp,z_+ +z_\perp]^{ab}F_-^{bi}(z_+,0,z_\perp)
\label{nonlocalforward}
\ega
where 
$$[x,y]\equiv {\rm Pexp}~ ig\!\int_0^1\! du ~(x-y)^\mu A_\mu(ux+(1-u)y)
$$ is a standard notation for straight-line gauge link
connection points $x$ and $y$.
It is easy to see that at even integer $j$ the expression in square brackets in Eq. (\ref{nonlocalforward})  is $\big({d\over dz'_+}\big)^{j-2}\delta(z'_+)$ so after
(j-2) integrations by parts we return to
``forward'' local operator (\ref{localforward}).

We will need also the light-ray operator integrated over $z_-$ direction
\bega
&&\hspace{-1mm}
\calf_j^{(-)}(z_\perp)~=~\int\! dz_- \!\int\! dz'_-\Big[{i\Gamma(j-1)\over 2\pi(z'_- +\ie)^{j-1}}+{i\Gamma(j-1)\over 2\pi(-z'_- +\ie)^{j-1}}\Big]
\nonumber\\
&&\hspace{-1mm}
\times~F^a_{+i}(z_- +z'_-,0,z_\perp)[z_- +z'_- +z_\perp,z_- +z_\perp]^{ab}F_+^{bi}(z_-,0,z_\perp)
\label{nonlocalminus}
\ega

Let us now demonstrate that the anomalous dimension of the LR operator (\ref{nonlocalforward}) is an analytic continuation of $\gamma_j$ of local operator (\ref{localforward}). First, let us rewrite the operator (\ref{nonlocalforward}) as follows
\bega
&&\hspace{-1mm}
\calf_j^{(+)}(z_\perp)
~=~{i\Gamma(j-1)\over \pi}\big(1-e^{-i\pi j}\big)\int_0^\infty\! dL_+~ L_+^{1-j}F(L_+,z_\perp)
\label{nonlocalforward1}
\ega
where
\begin{equation}
F(L_+,z_\perp)~=~\!\int\! dz_+~
F^a_{-i}(L_++z_+,0,z_\perp)[z_++z'_+ +z_\perp,z_+ +z_\perp]^{ab}F_-^{bi}(z_+,0,z_\perp)
\label{forwalr}
\end{equation}
is the ``forward'' gluon LR operator defined in Ref. \cite{Balitsky:1987bk}.

Evolution equation for the LR operator (\ref{forwalr})  has the form (see e.g.  Ref. \cite{Balitsky:1987bk})
\begin{eqnarray}
&&
\mu{d\over d\mu}F(L_+,x_\perp)~=~\!\int_0^1\! du ~K_{gg}(u,\alpha_s)F(uL_+,x_\perp)~~
\label{forwevoleq}
\end{eqnarray}
where  $K_{gg}(u,\alpha_s)$ is the DGLAP kernel in gluodynamics.  
The twist-2 operators (\ref{localforward}) are obtained by expansion of LR operator (\ref{forwalr}) in powers of $L_+$, 
so their anomalous dimensions are given by
\begin{eqnarray}
&&
\gamma_j(\alpha_s)
~=~-\!\int_0^1\! du ~u^{j-2}K_{gg}(u,\alpha_s) ~~~~~~\mu{d\over d\mu}\calo^{(+)}_j~=~-\gamma_j(\alpha_s)\calo^{(+)}_j
\end{eqnarray}

Now consider the anomalous dimensions of LR operator (\ref{nonlocalforward1}). Combining Eqs.  (\ref{forwalr}) and (\ref{forwevoleq}) one obtains
\begin{eqnarray}
&&\hspace{-1mm}
\mu{d\over d\mu}\calf_j(z_\perp)~
=~\!\int_0^1\!du~K_{gg}(u,\alpha_s)u^{j-2} \calf_j(z_\perp)
\label{gammacont}
\end{eqnarray}
Thus, we see that the above equation gives the explicit formula for analytic continuation of the anomalous dimensions to non-integer spins.

\bibliography{bibfail1}

\begin{thebibliography}{10}

\bibitem{Balitsky:1995ub}
I.~Balitsky.
\newblock {Operator expansion for high-energy scattering}.
\newblock {\em Nucl. Phys. B}, 463:99--160, 1996.

\bibitem{Kovchegov:1999yj}
Yuri~V. Kovchegov.
\newblock {Small x F(2) structure function of a nucleus including multiple
  pomeron exchanges}.
\newblock {\em Phys. Rev.}, D60:034008, 1999.

\bibitem{Balitsky:1987bk}
I.~I. Balitsky and Vladimir~M. Braun.
\newblock {Evolution Equations for QCD String Operators}.
\newblock {\em Nucl. Phys. B}, 311:541--584, 1989.

\bibitem{Lipatov:1985uk}
L.~N. Lipatov.
\newblock {The Bare Pomeron in Quantum Chromodynamics}.
\newblock {\em Sov. Phys. JETP}, 63:904--912, 1986.

\bibitem{Kotikov:2002ab}
A.~V. Kotikov and L.~N. Lipatov.
\newblock {DGLAP and BFKL equations in the $N=4$ supersymmetric gauge theory}.
\newblock {\em Nucl. Phys. B}, 661:19--61, 2003.
\newblock [Erratum: Nucl.Phys.B 685, 405--407 (2004)].

\bibitem{Kotikov:2010nd}
A.~V. Kotikov.
\newblock {The property of maximal transcendentality in the N=4 SYM}.
\newblock {\em Phys. Part. Nucl.}, 41:951--953, 2010.

\bibitem{Gromov:2015vua}
Nikolay Gromov, Fedor Levkovich-Maslyuk, and Grigory Sizov.
\newblock {Pomeron Eigenvalue at Three Loops in $\mathcal N=$ 4 Supersymmetric
  Yang-Mills Theory}.
\newblock {\em Phys. Rev. Lett.}, 115(25):251601, 2015.

\bibitem{Caron-Huot:2016tzz}
Simon Caron-Huot and Matti Herranen.
\newblock {High-energy evolution to three loops}.
\newblock {\em JHEP}, 02:058, 2018.

\bibitem{Velizhanin:2015xsa}
V.~N. Velizhanin.
\newblock {BFKL pomeron in the next-to-next-to-leading approximation in the
  planar N=4 SYM theory}.
\newblock 8 2015.

\bibitem{Wilson:1971dc}
Kenneth~G. Wilson and Michael~E. Fisher.
\newblock {Critical exponents in 3.99 dimensions}.
\newblock {\em Phys. Rev. Lett.}, 28:240--243, 1972.

\bibitem{Braun:2016qlg}
V.~M. Braun, A.~N. Manashov, S.~Moch, and M.~Strohmaier.
\newblock {Two-loop conformal generators for leading-twist operators in QCD}.
\newblock {\em JHEP}, 03:142, 2016.

\bibitem{Braun:2017cih}
V.~M. Braun, A.~N. Manashov, S.~Moch, and M.~Strohmaier.
\newblock {Three-loop evolution equation for flavor-nonsinglet operators in
  off-forward kinematics}.
\newblock {\em JHEP}, 06:037, 2017.

\bibitem{Ji:2023eni}
Yao Ji, Alexander Manashov, and Sven-Olaf Moch.
\newblock {Evolution kernels of twist-two operators}.
\newblock {\em Phys. Rev. D}, 108(5):054009, 2023.

\bibitem{Costa:2012cb}
Miguel~S. Costa, Vasco Goncalves, and Joao Penedones.
\newblock {Conformal Regge theory}.
\newblock {\em JHEP}, 12:091, 2012.

\bibitem{Costa:2013zra}
Miguel~S. Costa, James Drummond, Vasco Goncalves, and Joao Penedones.
\newblock {The role of leading twist operators in the Regge and Lorentzian OPE
  limits}.
\newblock {\em JHEP}, 04:094, 2014.

\bibitem{Costa:2023wfz}
Miguel~S. Costa, Vasco Goncalves, Aaditya Salgarkar, and Joao Vilas~Boas.
\newblock {Conformal multi-Regge theory}.
\newblock {\em JHEP}, 09:155, 2023.

\bibitem{Kravchuk:2018htv}
Petr Kravchuk and David Simmons-Duffin.
\newblock {Light-ray operators in conformal field theory}.
\newblock {\em JHEP}, 11:102, 2018.

\bibitem{Kologlu:2019mfz}
Murat Kologlu, Petr Kravchuk, David Simmons-Duffin, and Alexander Zhiboedov.
\newblock {The light-ray OPE and conformal colliders}.
\newblock {\em JHEP}, 01:128, 2021.

\bibitem{Chang:2020qpj}
Cyuan-Han Chang, Murat Kologlu, Petr Kravchuk, David Simmons-Duffin, and
  Alexander Zhiboedov.
\newblock {Transverse spin in the light-ray OPE}.
\newblock {\em JHEP}, 05:059, 2022.

\bibitem{Caron-Huot:2022eqs}
Simon Caron-Huot, Murat Kologlu, Petr Kravchuk, David Meltzer, and David
  Simmons-Duffin.
\newblock {Detectors in weakly-coupled field theories}.
\newblock {\em JHEP}, 04:014, 2023.

\bibitem{Jaroszewicz:1982gr}
T.~Jaroszewicz.
\newblock {Gluonic Regge Singularities and Anomalous Dimensions in QCD}.
\newblock {\em Phys. Lett. B}, 116:291--294, 1982.

\bibitem{Fadin:1998py}
Victor~S. Fadin and L.~N. Lipatov.
\newblock {BFKL pomeron in the next-to-leading approximation}.
\newblock {\em Phys. Lett. B}, 429:127--134, 1998.

\bibitem{Ciafaloni:1998gs}
Marcello Ciafaloni and Gianni Camici.
\newblock {Energy scale(s) and next-to-leading BFKL equation}.
\newblock {\em Phys. Lett. B}, 430:349--354, 1998.

\bibitem{Cornalba:2007fs}
Lorenzo Cornalba.
\newblock {Eikonal methods in AdS/CFT: Regge theory and multi-reggeon
  exchange}.
\newblock 10 2007.

\bibitem{Banks:1981nn}
Tom Banks and A.~Zaks.
\newblock {On the Phase Structure of Vector-Like Gauge Theories with Massless
  Fermions}.
\newblock {\em Nucl. Phys. B}, 196:189--204, 1982.

\bibitem{Mueller:1993rr}
Alfred~H. Mueller.
\newblock {Soft gluons in the infinite momentum wave function and the BFKL
  pomeron}.
\newblock {\em Nucl. Phys. B}, 415:373--385, 1994.

\bibitem{Mueller:1994jq}
Alfred~H. Mueller and Bimal Patel.
\newblock {Single and double BFKL pomeron exchange and a dipole picture of
  high-energy hard processes}.
\newblock {\em Nucl. Phys. B}, 425:471--488, 1994.

\bibitem{Balitsky:2007feb}
Ian Balitsky and Giovanni~A. Chirilli.
\newblock {Next-to-leading order evolution of color dipoles}.
\newblock {\em Phys. Rev. D}, 77:014019, 2008.

\bibitem{Balitsky:2009xg}
Ian Balitsky and Giovanni~A. Chirilli.
\newblock {NLO evolution of color dipoles in N=4 SYM}.
\newblock {\em Nucl.\ Phys.\ B}, 822:45--87, 2009.

\bibitem{Braun:2013tva}
V.~M. Braun and A.~N. Manashov.
\newblock {Evolution equations beyond one loop from conformal symmetry}.
\newblock {\em Eur. Phys. J. C}, 73:2544, 2013.

\bibitem{Balitsky:2013npa}
Ian Balitsky, Vladimir Kazakov, and Evgeny Sobko.
\newblock {Two-point correlator of twist-2 light-ray operators in N=4 SYM in
  BFKL approximation}.
\newblock {\em Nucl. Phys. B}, 993:116267, 2023.

\bibitem{Balitsky:2014sqe}
Ian Balitsky.
\newblock {NLO BFKL and anomalous dimensions of light-ray operators}.
\newblock {\em Int. J. Mod. Phys. Conf. Ser.}, 25:1460024, 2014.

\bibitem{Cornalba:2008qf}
Lorenzo Cornalba, Miguel~S. Costa, and Joao Penedones.
\newblock {Eikonal Methods in AdS/CFT: BFKL Pomeron at Weak Coupling}.
\newblock {\em JHEP}, 06:048, 2008.

\bibitem{Balitsky:2009yp}
Ian Balitsky and Giovanni~A. Chirilli.
\newblock {High-energy amplitudes in N=4 SYM in the next-to-leading order}.
\newblock {\em Phys. Lett.}, B687:204--213, 2010.

\bibitem{Balitsky:2018irv}
Ian Balitsky.
\newblock {Structure constants of twist-two light-ray operators in the triple
  Regge limit}.
\newblock {\em JHEP}, 04:042, 2019.

\bibitem{Caron-Huot:2015bja}
Simon Caron-Huot.
\newblock {Resummation of non-global logarithms and the BFKL equation}.
\newblock {\em JHEP}, 03:036, 2018.

\bibitem{Balitsky:2012bs}
Ian Balitsky and Giovanni~A. Chirilli.
\newblock {Photon impact factor and $k_T$-factorization for DIS in the
  next-to-leading order}.
\newblock {\em Phys. Rev. D}, 87(1):014013, 2013.

\end{thebibliography}
\bibliographystyle{JHEP}

\end{document}